# Boron Nitride Coatings for the Enhanced Detection of Neutrons in CR-39

Noah D'Amico [a]*, Sandeep Puri [a], Ian Jones [a], Andrew Gillespie [a], Cuikun Lin [a], Bo Zhao [b], R. V. Duncan [a]

[a] *Center for Emerging Energy Sciences, Department of Physics and Astronomy, Texas Tech University, Lubbock, Texas, USA*
[b] *College of Arts and Sciences Microscopy, Texas Tech University, Lubbock, Texas, USA*
**\* Corresponding Authors:** ndamico@ttu.edu

**Keywords:** Thin Film, Nuclear Detection, Plastic Nuclear Track Detector, AmBe, $^{252}$Cf, 14 MeV Neutrons

## Abstract

The neutron detection efficiency of Columbia Resin 39 (CR-39) nuclear track detectors was assessed for AmBe, $^{252}$Cf, and D-T (14 MeV) neutron source spectra. A boron nitride (BN) coating for CR-39 was developed to enhance the neutron detection efficiency by converting neutrons into energetic alpha particles through the well-known $^{10}$B(n,α)$^{7}$Li reaction. Separate partially coated CR-39 pieces were exposed to each neutron source and subsequently analyzed under optical microscope and through large-area Scanning Electron Microscopy (SEM) imaging over the irradiated area. The detection efficiency (tracks per neutron) was evaluated for each source spectra under optical and scanning electron microscopes and with or without BN coating. This resulted in a comprehensive guide to neutron detection with various sources using CR-39.

## 1.0 Introduction

Columbia Resin 39 (CR-39) is a high-quality plastic developed for use as a Plastic Nuclear Track Detector (PNTD). Energetic ions incident on the surface of CR-39 leave defined tracks as they are slowed down in the material, and these tracks may be enhanced by chemical etching, making them easily visible under an optical microscope. The nature of this detector allows for high accuracy in absolute particle counts, as there is a one-to-one correlation between incident ions above a threshold energy and surface tracks.[1–8]

A primary challenge in working with CR-39 is the lack of sensitivity to neutrons, which do not have the same track-creating mechanism as charged particles.[9–12] Typically, energetic neutrons will leave tracks in CR-39 through elastic scattering with the atoms of the CR-39, with the probability for a neutron to leave a track depending directly upon the elastic scattering cross section with the nuclei in the CR-39. Additionally, neutron capture reactions with the constituent atoms of CR-39 ($C_{12}H_{18}O_{7}$) can result in the creation of energetic alpha particles or protons, also with low cross sections. Thermal neutrons, on the other hand, cannot impart sufficient energy to atoms to create visible tracks in the CR-39. Past experiments have been attempted to solve this lack of thermal neutron detection, including baking boron into the CR-39 plastic to take advantage of its high thermal neutron capture cross section.[13] The problem with this is that the etching process must be repeated over and over until the entire CR-39 sample is analyzed, as the neutron reactions are spread throughout the volume of the plastic.

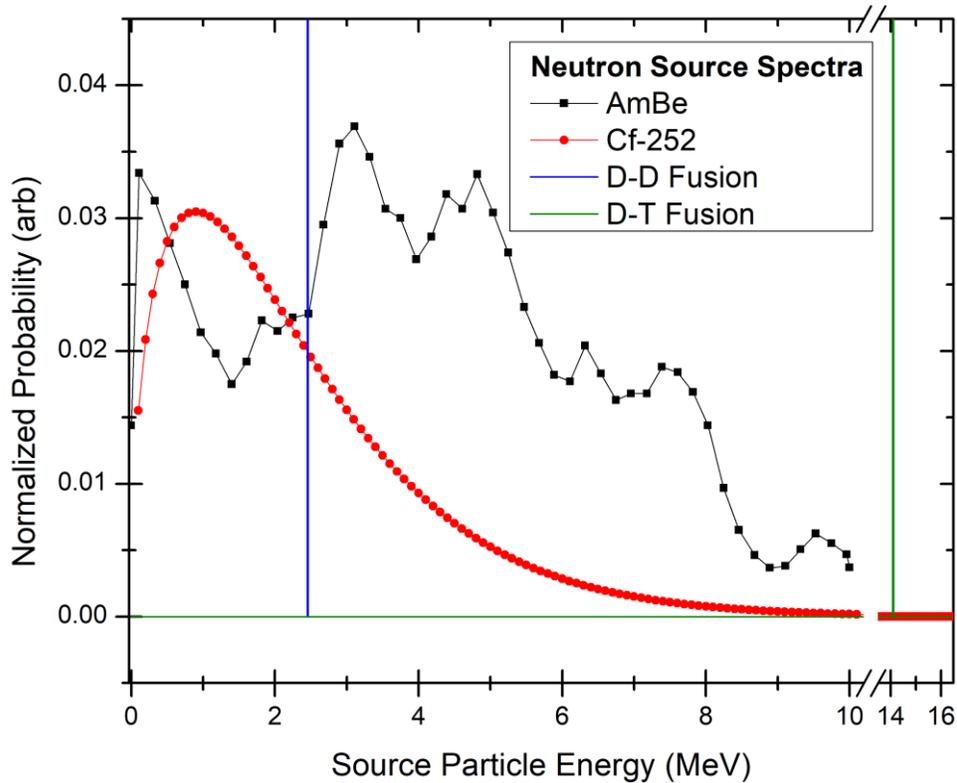

*Figure 1*: Neutron spectra for various sources.

We have created a boron-rich coating on top of CR-39 that enhances its neutron detectivity for various neutron source spectra (shown in **Figure 1**) including AmBe, $^{252}$Cf, and D-T fusion and only requires one round of etching, as all the tracks originate at the coating on the surface. The mechanism of boron thermal neutron capture allows for a high efficiency of thermal neutron conversion into energetic alpha particles. The alpha particles can then exit the thin film of BN and hit the CR-39, creating a track that is clearly visible after etching. While the cross section of this (n,α) reaction decreases as neutron energy increases, experimental evidence shows clear increases of up to a factor of 10x neutron detection in CR-39 alone, even at neutron energies up to 5 MeV. At 5 MeV, the cross sections of alpha creation within the CR-39 constituent atoms (such as the $^{16}$O(n,α)$^{13}$C reaction) become higher than that of boron-10. Additionally, elastic collisions with neutrons of energies above 5 MeV began to impart sufficient energy (generally considered ~500 keV for heavy atoms[7,14]) for carbon and oxygen ions to leave tracks in the CR-39. Finally, the well-known $^{12}$C(n,3α)n reaction occurs within CR-39 with neutrons above 8 MeV[15], though no detection algorithms for triple alpha tracks were included in the analysis for this paper.

$^{10}$B + n → $^{4}$He + $^{7}$Li + 2.31 MeV (σ = ~3800 barns at 0.025 eV)[1]

It is worth noting that there is also a nitrogen thermal neutron capture

$$^{14}N + n \rightarrow {}^{1}H + {}^{14}C + 0.63 \text{ MeV} (\sigma = \sim1.82 \text{ barns at } 0.025 \text{ eV})$$

that does not appear to have a noticeable effect on CR-39 when compared to the boron component, given its 3 orders of magnitude smaller cross section, and the fact that this cross section falls off at a similar rate at higher energies. This BN coating can be applied to one side of the CR-39 to give increased neutron efficiency, while keeping the other side uncoated for typical energetic ion analysis, creating a dual measurement system from the cost-effective plastic. This paper explores the enhancements to neutron detection made by BN coatings, as well as the experimental detection efficiency of uncoated CR-39 to AmBe, $^{252}$Cf and D-T neutrons with CR-39 etching conditions of 12.5 M NaOH at 80°C for one hour.

## 2.0 Experiment: Coating, Exposure, Simulation, SEM, Analysis
### 2.1 Coating

To create the boron nitride (BN) coating mixture, 1 gram of BN powder (-325 mesh) was measured in a measuring boat. 1 gram of 'Elmer's School Glue', which is polyvinyl acetate (PVA) and water, was slowly added to the powder. 1.2 mL of water was added to the mixture. The mixture was stirred carefully until visibly homogenous (3 minutes) then transferred to a mortar and pestle for stronger grinding. The mixture was ground for 20 minutes, resulting in a homogenous paste with no visible BN particulates remaining.

A VTC-100A spin coater from MTI Corporation[16] was used to coat the CR-39 with BN glue. A 1 square inch sample of CR-39 was placed onto the spin coater with a vacuum seal, and a 5-mm diameter drop of the BN paste was placed in the middle of the CR-39. The spin coater was then immediately ramped up to 1500 rpm and run for 7 seconds, and then immediately ramped to 3500 rpm and run for 12 seconds. After spinning stopped, the coating was allowed 10 minutes to ensure that it fully dried. This creates a BN-glue layer with an average thickness of 10 microns. Note that this coating is robust enough to be stored loosely in a bag with similar CR-39 pieces, but strong contact and contact with water will begin to remove the coating.

### 2.2 AmBe Neutron Exposure

A 10 mCi (2x10$^4$ neutrons/s) AmBe source was used to test the detection efficiency of AmBe spectrum neutrons within CR-39. The source was cylinder with a diameter and height of 1 cm. A 1-cm-thick coaxial cylindrical shell of polyethylene was used as a moderator. A BN-coated (one side, other side uncoated) CR-39 of 3-mm thickness was placed on the side of the cylinder with the BN side facing away from the source. This configuration, shown in **Figure 2**, ensures that the thickness of the thin film will not hinder the detectivity, as a very thick film would likely create most neutron captures on the side nearest the source, not allowing any ions (with a penetration depth of 8-20 microns) to make it to the CR-39 surface. With the source facing the uncoated side of the CR-39, neutrons will be incident on the inside of the coating after traveling through the CR-

39, allowing roughly half of the alpha particles created by boron neutron capture to create tracks on the CR-39.

The CR-39 was irradiated for 32 days under these conditions. After the irradiation, the coating was cleaned off with water, and the CR-39 was etched in an 80 °C, 6.25 M NaOH bath for 1 hour with a stir rod at a constant stirring rate of 300 rpm to ensure even etching. The CR-39 was then analyzed under both an optical microscope and SEM. A 10x increase in neutron detection efficiency was found on the boron-coated side of the CR-39 with this neutron energy.

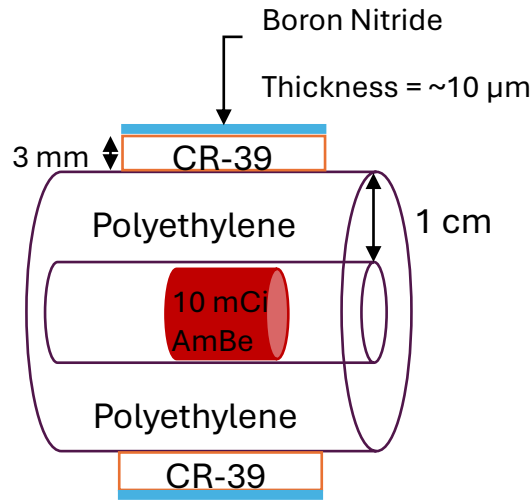

*Figure 2:* Experimental setup for thermal neutron exposure of BN-coated CR-39. The uncoated front side of the CR-39 serves as an inherent control to establish the detection enhancement of the BN layer. MCNP simulation was used to calculate the expected total neutron flux on the surfaces of the CR-39 from the exposure.

### 2.3 $^{252}$Cf Neutron Exposure

A 0.75 mCi $^{252}$Cf stainless steel sealed source (~3.3 x $10^6$ neutrons/s) was placed directly under a piece of BN-coated CR-39, with the BN coating facing away from the source. This was exposed for 11 days. Upon halting the irradiation, the etching and imaging procedure discussed above was used. A 4x increase in neutron detection efficiency was found on the boron-coated side of the CR-39 with this neutron energy.

### 2.4 D-T Fusion Neutron Exposure

A BN-coated piece of CR-39 was placed directly on the target plane of a P383 D-T neutron[17] generator running at 120kV and 70μA. This produced neurons at a rate of 4 x $10^8$ neutrons/s, and it was operated for 2 hours. Upon halting the irradiation, the etching and imaging procedure discussed above was used. No visible difference was found between the boron-coated and uncoated sides at this high neutron energy.

### 2.5 Simulations

A volume source of AmBe[18], inside polyethylene was simulated with Monte Carlo n-Particle (MCNP[19]) transport calculations using the same dimensions as shown in **Figure 2**. A 1/8"-thick slab of CR-39 with a 10-micron layer of Boron Nitride mixed with Elmer's Glue was located above this source. This BN layer material definition assumes a 50% mixture of PVA with BN using atomic ratios. The CR-39 has been split into three layers for the purpose of this simulation. All stacked layers are 2.5 cm x 2.5 cm wide.

These simulations were performed to determine the total number of neutrons hitting the CR-39 on both the front and back sides. We're only interested in the solid angle flux, minus the neutrons that may have reacted within the moderation or otherwise don't make it to the sample. These results will be compared to the tracks detected by AI analysis to obtain the efficiency. Additionally, these simulations were used to obtain an energy distribution of the neutrons when they hit the CR39 or the BN coating. The total 1/8"-thick slab of CR-39 was separated into three layers to determine differences in neutron counts in the front-most layer compared to the back-most layer. The closest and farthest layers from the source were 0.1 mm thick and the middle layer was 2.975-mm thick. Simulations were used to determine a total neutron flux so that a "neutron detection efficiency" using the number of tracks observed on the boronated side divided by the total amount of incident neutrons. All simulations were run using $10^9$ particle histories (shown in **Table 1**) and results were scaled to the source activity of $2.2 \times 10^4$ n/s.

| Layer Name | # Neutron Tracks Entering | % of Source |
|---|---|---|
| Boron Nitride | 97,740,736 | 9.77% |
| CR-39 Farthest Layer | 98,613,028 | 9.86% |
| CR-39 Closest Layer | 131,884,826 | 13.19% |

*Table 1*: Neutron tracks entering each simulated layer of CR-39 in AmBe irradiation experiment with $10^9$ histories.

The rate of neutrons entering CR-39 is about $2.901 \times 10^3$ n/s or $2.507 \times 10^8$ n/day. The rate of neutrons entering BN is about $2.150 \times 10^3$ n/s or $1.858 \times 10^8$ n/day. In a 32-day exposure, this results in $8.001 \times 10^9$ neutrons incident on the closest layer (uncoated) and $5.945 \times 10^9$ on the farthest layer (BN).

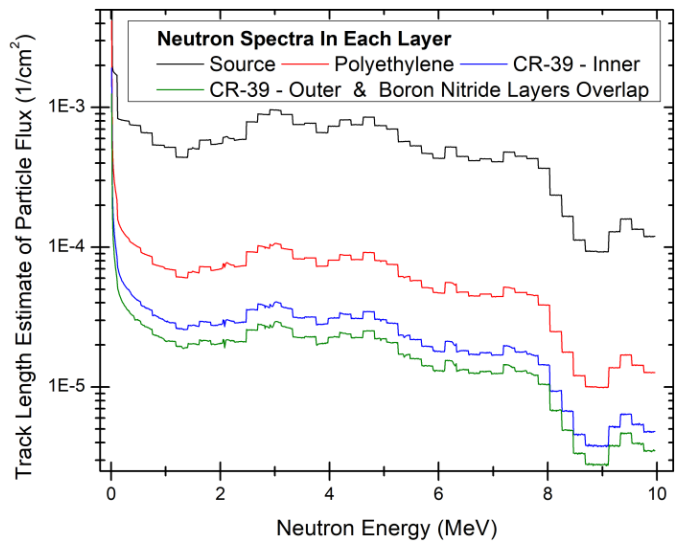

*Figure 3*: Neutron energy spectra in various layers of CR-39 and BN coating.

There was not significant thermalization of the AmBe source neutron spectrum within each material, as shown in **Figure 3**, where the relative neutron energy spectrum remains the same, and only total particle count is decreased by the polyethylene layer. The differences in the spectra heights are mainly due to different solid angles being subtended by each layer. Some attenuation reduced the flux in successive layers.

A second set of simulations was performed to determine the extent that source neutrons could be thermalized by the polyethylene moderator. A surface source of neutrons was incident on polyethylene of variable thickness. This was performed to determine whether sufficient thermalization could be achieved without reducing the flux to an unacceptable level. Thermalizing the neutrons will enhance the neutron capture by $^{10}$B. The overall flux and the changes to the lowest energy bin were tracked with increasing polyethylene thickness. As expected, the attenuation of neutrons increases with polyethylene thickness. This does not result in significant thermalization. The flux for neutrons of 0.001-MeV and lower energies only decreases with increasing polyethylene thickness. This means that there is no benefit in including large moderators for experiments with the AmBe source.

The experimental conditions of the $^{252}$Cf exposure were simulated in MCNP using $10^9$ histories to determine the expected incident neutrons on the uncoated and BN-coated faces of CR-39 using 20-micron layers. Simulations predicted that the uncoated layer would receive 8.48% of the total neutron output and the BN layer would receive 7.06%. For a $3.3 \times 10^6$ n/s source activity, this corresponds to $2.17 \times 10^{12}$ and $1.81 \times 10^{12}$ total incident neutrons respectively for the 11-day experiment.

The experimental conditions of the D-T neutron irradiation were also simulated in MCNP using $10^9$ histories to determine the expected incident neutrons on the uncoated and BN-coated faces of CR-39 using 20-micron layers. The flux distribution of a ThermoFisher P385 neutron generator[20] was used for simulation, which may differ slightly from the P383 used for the physical experiment – note that neither are isotropic sources. Simulations predicted that the CR-39 detector would receive about 2.05% of the total neutron output and that about $5.9 \times 10^{10}$ total neutrons would be incident on each side of the CR-39 for the 2-hour irradiation. There is little difference between the total flux of the two sides of CR-39 because the thickness of the CR-39 is negligible with respect to the neutron generator dimensions in this experiment.

## 2.6 SEM and Large Area Mapping

SEM was used to create large area maps to analyze neutron-induced tracks over a wide area using a Zeiss Crossbeam 540 with Atlas 5 large area mapping imaging system. This SEM is a powerful tool for advanced materials research capable of precise SEM imaging to enable high-resolution surface imaging and site-specific sample preparation. With as low as 5-nanometer resolution, it can deliver high-contrast imaging of surfaces and subsurface structures. Its high throughput capabilities were combined with advanced automation methods to optimize fast data acquisition with our automated workflow method. The images were collected under 5 kV accelerating voltage with an SE detector. The large area mapping was set at 200 nm resolution with 6.4 µs dwell time, 600 µm x 600 µm tile size and 10% overlap. It should be noted that this overlap setting is purely for the automated construction of the mosaic and does not influence the source images. Therefore, the total particle count is not affected by the 10% overlap.

Images were captured robotically, spanning the entire surface, and stitched together using Mosaic software[21]. The AI system was trained to characterize particle tracks left on the surface of the CR-39 large area map. The images that constitute the large area map were analyzed by the AI individually and later stitched together to recreate a post-processing large area map.

### 2.7 AI Particle Detection Analysis

MATLAB machine learning and object detection packages were used to develop an AI object detection program for use on CR-39. For automated analysis of particle tracks, a transfer-learning-based Region-based Convolutional Neural Network (R-CNN) was employed.[22–26] The R-CNN algorithm utilizes a pre-trained deep residual neural network (ResNet-101) backbone to identify and classify regions of interest, specifically tailored to detect the small, elliptical features characteristic of alpha particle tracks. To improve detection accuracy and computational efficiency, a Canny edge detection algorithm was applied to highlight particle tracks in the SEM images. For more details on this AI CR-39 track detection program, see D'Amico, *et al*. (2025).[27]

Two detection schemes were used for the $^{252}$Cf analysis. One aimed to detect all possible tracks made in the CR-39, and used Canny parameters (0.1, 0.2, 1.4), filtering boxes with areas above 1000 square pixels, then combined these results with Canny parameters (0.15, 0.35, 4), eliminating bounding boxes from the lower threshold Canny detection if they had over 20% overlap. This detection scheme counted small tracks accurately while also getting the correct size of larger, clearer tracks. The second scheme employed was meant to detect only tracks that would be visible under an optical microscope without the resolution to see the micron diameters and lower contrast of (relatively low energy) neutron recoils. This scheme used only the higher threshold Canny parameters (0.15, 0.35, 4) and filtered out all boxes smaller than 800 square pixels. The results of this second scheme were compared with analysis of optical images (by hand) to ensure accuracy in BN coating detection enhancement values.

### 3.0 Results and Discussion

The BN-glue coating process creates uniform coatings of 10 ± 2-micron thickness across 6.25 cm$^2$ pieces of CR-39. SEM imaging of the edge and surface of the coating can be seen in **Figure 4.**

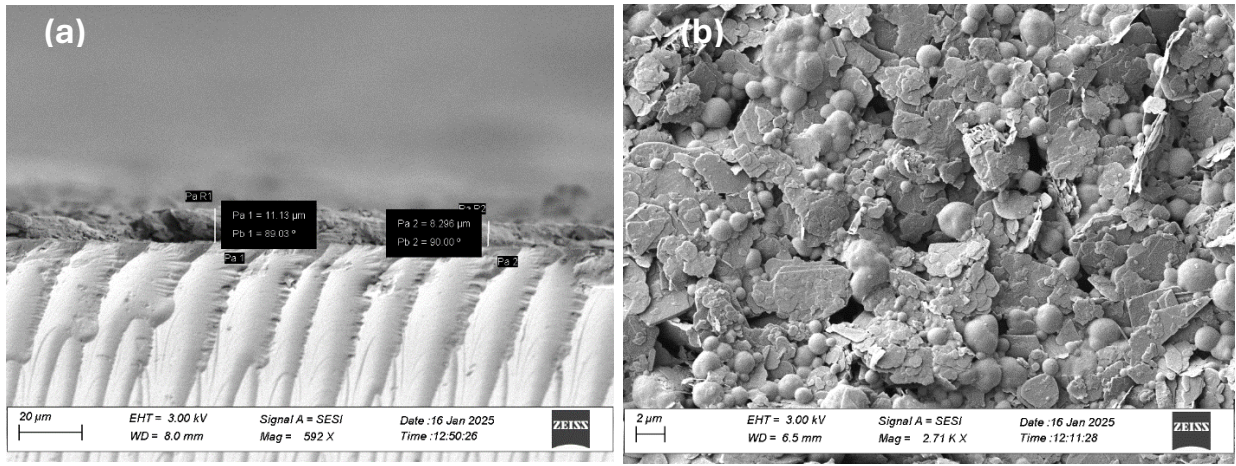

*Figure 4:* **(a)** SEM cross section of CR-39 with surface coating, demonstrating thickness of 10 ± 2 μm, and **(b)** surface of coating, where the crystalline plates are BN crystal, and the amorphous structures are PVA.

The detection enhancements made by the BN coatings must be analyzed while noting that SEM imaging allows for much greater resolution and visualization of neutron-induced tracks in CR-39 than optical imaging. CR-39 records many small proton recoils that are too small to detect (or distinguish from surface defects) under optical microscope, but appear quite clear and can easily be counted under SEM. Thus, there is a different overall detection enhancement factor depending on the imaging method used. For example, the $^{252}$Cf sample exhibits no increase in total tracks from BN coating when SEM images are analyzed using AI (801151 tracks on BN side, 822403 tracks on uncoated side) because all small proton recoil tracks can be easily seen under SEM. However, a 4.4x increase in visible tracks under a high-end optical microscope is found (average of 110 tracks on BN side, 25 tracks on uncoated side per 280x280 micron image), where small proton recoil tracks are not visible, so (n,α) reactions are relied upon for detection. This matched the SEM AI detection algorithm meant to mimic required conditions for detection in *our* optical microscope, which also showed a 4x increase (48588 tracks on BN side, 12893 tracks on uncoated side). From these results, it can be concluded that SEM imaging allows for a 1.5x10$^{-6}$ efficiency (tracks per neutron incident) of $^{252}$Cf neutron detection with CR-39, with or without a BN layer. Under high-power optical imaging, the CR-39 has a low natural efficiency of 2.4x10$^{-8}$, which can be increased to 10$^{-7}$ with a 10-micron BN layer.

**Figure 5** shows the difference between $^{252}$Cf tracks from coated and uncoated CR-39 under optical and SEM microscopes, as well as the different AI detection schemes employed to quantify the increase in detection efficiency for both microscopy methods. These images make it clear that high-resolution microscopy is required to accurately detect ~1 MeV level neutron dosage using CR-39. Under an optical microscope, it is impossible to distinguish micron-sized neutron recoil tracks from surface impurities on the CR-39, making the neutron dose count on an uncoated piece of CR-39 very subjective; without nanometer-scale resolution, coating enhancements allow for more accurate measurement.

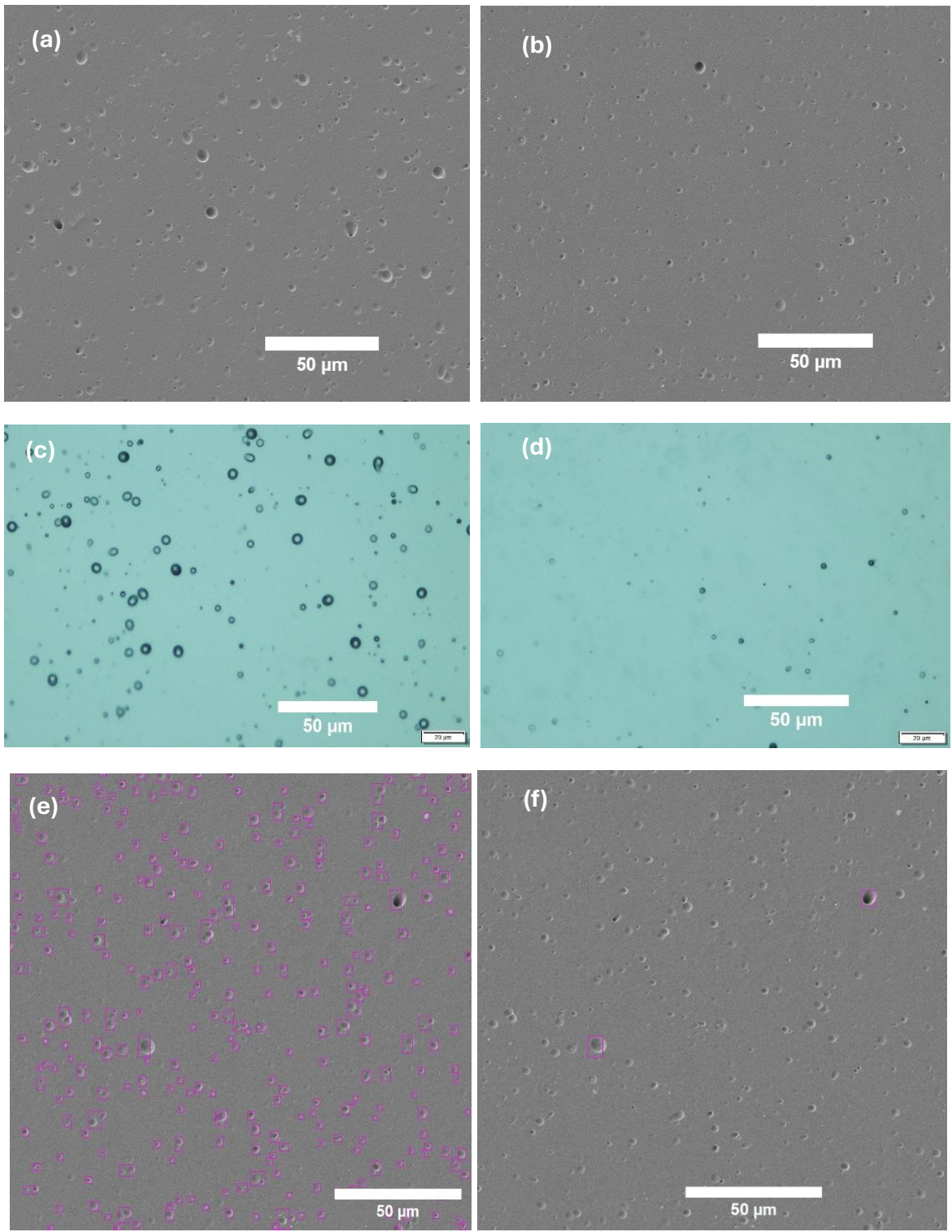

*Figure 5*: (a) $^{252}$Cf neutron tracks on BN-coated side of CR-39 under SEM. (b) $^{252}$Cf neutron tracks on uncoated side of CR-39 under SEM. (c) $^{252}$Cf neutron tracks on BN-coated side of CR-39 under optical microscope. (d) $^{252}$Cf neutron tracks on uncoated side of CR-39 under optical microscope. (e) First detection algorithm applied to count all tracks visible on

CR-39 surface from SEM. (f) Second detection algorithm applied to count only tracks that would be visible on surface using optical microscope.

The effect of BN on neutron detectability from a sealed AmBe source was studied using the same AI detection techniques. There was a much clearer enhancement of tracks under optical microscope when compared to $^{252}$Cf (see **Figure 6**). Using the same Canny parameters as before, the BN was found to create a 10x increase in tracks visible using an optical microscope (74981 on BN side, 7523 on uncoated side). Under SEM, 321950 tracks were detected on the BN side, and 211476 tracks were detected on the uncoated side. Considering the difference in solid angle that each side subtended with respect to the AmBe source, this represents a 2.2x enhancement in neutron detection efficiency under SEM. From these results, it can be concluded that SEM imaging allows for a $2.5 \times 10^{-5}$ natural efficiency of AmBe neutron detection with CR-39, and $5.5 \times 10^{-5}$ with a BN layer. Under high-power optical imaging, the CR-39 has a low natural efficiency of $10^{-6}$, which can be increased to $10^{-5}$ with a 10-micron BN layer when integrated over the full neutron energy spectrum of the AmBe source.

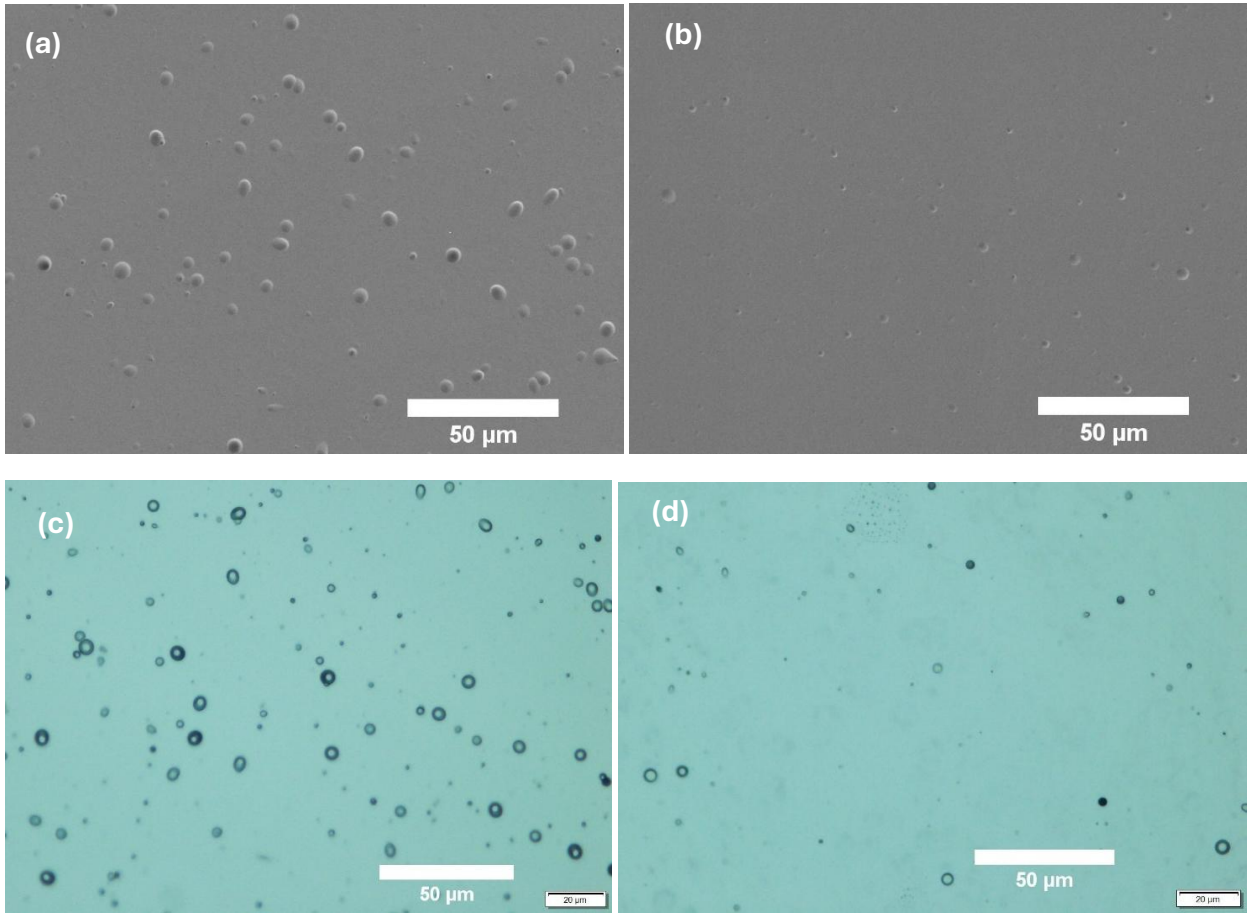

*Figure 6*: (a) AmBe neutron tracks on BN-coated side of CR-39 under SEM. (b) AmBe neutron tracks on uncoated side of CR-39 under SEM. (c) AmBe neutron tracks on BN-coated

side of CR-39 under optical microscope. (d) AmBe neutron tracks on uncoated side of CR-39 under optical microscope.

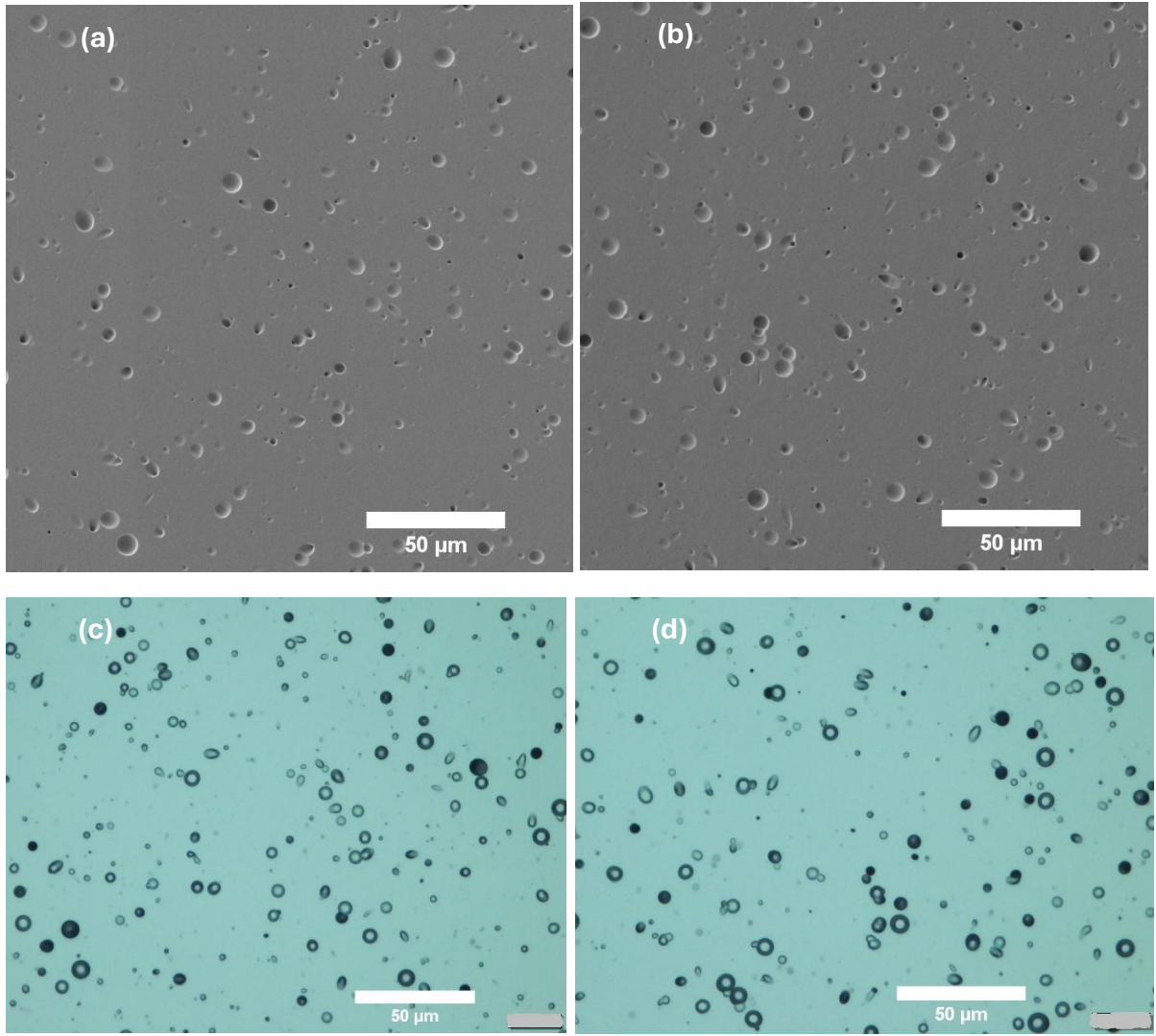

*Figure 7*: (a) D-T neutron tracks on BN-coated side of CR-39 under SEM. (b) D-T neutron tracks on uncoated side of CR-39 under SEM. (c) D-T neutron tracks on BN-coated side of CR-39 under optical microscope. (d) D-T neutron tracks on uncoated side of CR-39 under optical microscope.

The BN coating has no statistically significant effect on the detection of 14 MeV neutrons using CR-39 under SEM, as seen in **Figure 7** (633201 tracks on BN side, 612995 on uncoated side). The detection efficiency of CR-39 with 14 MeV neutrons under SEM was found to be $4.3 \times 10^{-5}$. Unlike both AmBe and $^{252}$Cf, no BN detection enhancement is found under optical microscope either (238415 and 241944, respectively). This is to be expected, as the cross section of the $^{10}$B(n,α)$^7$Li is significantly smaller than neutron recoil cross sections at this energy. The total number of tracks found under optical microscope is much closer to the number found using SEM when analyzing 14 MeV neutron irradiation. This is likely because the mechanisms for track creation change. With lower (<5 MeV) energy neutrons, elastic collisions can only create tracks from protons. Elastic collisions with carbon and oxygen atoms within the plastic transfer a maximum of 400-500 keV (though often much lower), which is generally thought to be too low for a heavy ion to leave a significant track after etching. At these neutron energy levels, (n,α) reactions dominate the creation of large tracks visible under optical microscope. In contrast, at 14 MeV neutron energy the cross sections for (n,α) reactions decrease, but elastic scattering of carbon and oxygen become viable sources of large tracks in CR-39, as their energies can exceed 1 MeV after collisions. These tracks are larger because the ions have high positive charge and therefore deposit energy more rapidly, creating a wider range of broken bonds around 7-10 microns in diameter. Larges tracks like these can be distinguished from surface defects in the plastic much easier than small proton recoil tracks under optical microscope, hence creating a higher (and lower uncertainty) detection efficiency.

In the case of both AmBe and $^{252}$Cf, track size appears to be a limiting factor for detection, especially under optical microscope. When tracks are too small to be differentiated from dust or surface defects of the plastic, they are not counted towards the total track value. Some of these ignored dots are later shown to be true tracks when analyzed under SEM microscope. We recognize that longer or more intense etching processes could help further develop these tracks, enabling them to be distinguished more easily, therefore increasing the detection efficiency of lower energy neutrons. One concern with increasing etching time or intensity is the further increase in size of larger tracks made by alpha particles or heavy ion recoil within the CR-39. We have found that doubling etch time of our solution leads to a decrease in total particle detections in some cases due to overwhelming overlap of large tracks at the exposure levels and exposure times stated above.

It is also important to note that the greatest theoretical enhancement to CR-39 neutron detection using BN coating exists when the incident neutrons have thermal energies. At these energies, uncoated CR-39 has no mechanisms for track creation, so thermal neutrons are undetectable. In contrast, the $^{10}$B neutron capture cross section is highest at thermal neutron energies. This study demonstrates the efficacy of BN coating at higher neutron energies. It is clear that the significantly greater cross section at thermal neutron energies (~$10^3$ barns vs ~1 barn) will lead to an even greater detection enhancement.

**4.0 Conclusion**

CR-39 detectors were exposed to AmBe, $^{252}$Cf, and D-T fusion neutron sources to determine the neutron detection efficiency of CR-39 at various energy spectra. Boron nitride coatings for CR-39 were developed for enhancement of neutron detection through the $^{10}$B(n,α)$^7$Li reaction, which resulted in the creation of energetic alpha particles that readily leave tracks in CR-39. An AI-based analysis tool was used to give accurate, automated counts of tracks on each piece of CR-39. The boron nitride coating was found to create large enhancements in neutron detection efficiency under optical microscope. Neutron detection efficiency was increased by a factor of 4x for the $^{252}$Cf source and by a factor of 10x for the AmBe source. No statistically significant enhancement was observed for detecting very high energy (14 MeV) neutrons from the D-T generator. Using SEM, only AmBe neutron detectivity was increased by a factor of 2.2x while using BN coatings. It was also discovered that SEM imaging allows for 1-2 orders of magnitude higher detection efficiency than optical imaging in below 14-MeV neutrons, as the higher resolution can more readily distinguish proton recoil tracks from micro-scale surface defects or dust.

## 5.0 Conflict of Interest

The authors declare that the research was conducted in the absence of any commercial or financial relationships that could be construed as a potential conflict of interest. One author (RVD) has a financial interest in a new Texas Tech University spin-off company named BlankSlate Innovation, LLC (BSI) that is in the process of licensing back patent rights from Texas Tech University on this CR-39 detection technology, but this arrangement is in its formative stages, and the commercial operations of this company did not bias or otherwise influence this scientific study in any way.

## 6.0 Author Contributions

All authors contributed to the conception and design of the research reported here. ND wrote the first draft of the manuscript. All authors contributed to manuscript revision, read, and approved the submitted version.

## 7.0 Funding

This work was supported by the Department of Energy award No. DE-AR0001736.